\documentclass[a4paper,12pt]{article}
\usepackage{epsfig}
\usepackage{citesort}
\date{\today}
 \normalsize

\newcommand{\bmat}{\left(\begin{array}}
\newcommand{\emat}{\end{array}\right)}
\newcommand{\be}{\begin{equation}}
\newcommand{\ee}{\end{equation}}
\newcommand{\bea}{\begin{eqnarray}}
\newcommand{\eea}{\end{eqnarray}}

\def\lsim{\raise0.3ex\hbox{$\;<$\kern-0.75em\raise-1.1ex\hbox{$\sim\;$}}}
\def\gsim{\raise0.3ex\hbox{$\;>$\kern-0.75em\raise-1.1ex\hbox{$\sim\;$}}}
\def\Frac#1#2{\frac{\displaystyle{#1}}{\displaystyle{#2}}}

\begin{document}
\renewcommand{\thefootnote}{\fnsymbol{footnote}}
\rightline{IPPP-04-07} \rightline{DCPT-04-14} 
\vspace{.3cm}
{\Large
\begin{center}
{\bf Flavor violation and extra dimensions}
\end{center}}
\vspace{.3cm}

\begin{center}
{\bf S. Khalil$^{1,2}$ and R. Mohapatra$^{3}$}\\
\vspace{.3cm}
$^1$ \emph{IPPP, Physics Department, Durham University, DH1 3LE,
Durham,~~U.~K.}
\\
$^2$ \emph{Ain Shams University, Faculty of Science, Cairo, 11566,
Egypt.}
\\
$^3$ \emph{Department of Physics, University of Maryland, College Park, MD 20742, USA.}
\end{center}

\vspace{.3cm}
\hrule \vskip 0.3cm
\begin{center}
\small{\bf Abstract}\\[3mm]
\end{center}
 We analyze new sources of flavor violation in models with extra
dimensions. We focus on 
three major classes of five dimensional models: models with universal
extra dimension, 
models with split fermions, and models with warped extra dimension. 
We study the implications of these new sources on the associate CP
violating asymmetries to the 
rare $B$-decays. We show that among these models only the split fermions
scenario may accommodate 
the recent experimental deviation between the CP asymmetry of $B_d \to \phi
K_S$ and $\sin 2 \beta$.
 
\begin{minipage}[h]{14.0cm}
\end{minipage}
\vskip 0.3cm \hrule \vskip 0.5cm
%
\section{\large{\bf{ Introduction}}}
It is a common belief held
by many theorists for a long time that our world is 
not limited to just four dimensions but it can extend to as many as
eleven. More recently, extra dimensions have been proposed as an
alternative way to address the origin
of the TeV scale \cite{extra1,extra2,extra3}. This has led to a new surge
of activities to look for their implications for particle phenomenology.
 
There are many variations on the basic theme of extra
dimensions in order to discuss different puzzles of the standard
model; for instance, there have been attempts to
understand geometrically the origin of the fermion mass hierarchy
\cite{Arkani-Hamed:1999dc,Gherghetta:2000qt}, CP violation \cite{chang}
etc. A generic feature of this class of models is that they
contain new sources of flavor violation, due to the coupling of zero-mode
fermions to the 
Kaluza-Klien (KK) excitations of the gauge and Higgs bosons, which might
imply severe constraints on 
the string scale.  This raises the possibility of a 
conflict between the new 
physics scale preferred by the solution of the hierarchy problem ($M_S  
\sim  \mathcal{O}(1)$ TeV)
and the one needed to satisfy the flavor constraints. Thus the issue of
flavor violation in extra dimension models is an important one and has
only recently begun to be addressed. In particular,
in several recent papers \cite{Buras:2003mk} this has been discussed in
the context of the universal extra dimension models
\cite{Appelquist:2000nn}. Our goal in this
paper is to pursue this in the context of different extra dimension
models and to isolate possible flavor violation signals.

We are specially motivated by the recent experimental measurements
\cite{Belle3,BaBar3} of the CP asymmetries of $B_d \to \phi K_S$ 
($S_{\phi K_S}$) and $B_d \to \eta' K_S$ ($S_{\eta' K_S}$), which seem to
indicate a deviation from standard model predictions by
more than two standard deviations. This has been 
considered as perhaps a hint for new CP and/or flavor violating sources
beyond the SM. 
There have been attempts to interpret this within the framework of 
supersymmetric theories with a new flavor structure beyond the Yukawa 
matrices i.e. in the squark mass matrices.
These results can be easily accommodated \cite{KK-phiks,phiks-susy} even
if the 
phase in the CKM mixing matrix is the only source of CP violation
\cite{veronica}.  
If this anomaly is confirmed it would be natural to look for the type of 
new physics that can generate such deviation between $S_{\phi K_S}$ and
$S_{J/\psi K_S}$ and to ask if models with extra dimensions can accommodate 
this result.

In this paper we study the implications of models with extra dimensions on
these processes. We analyze three major classes of five dimensional models. 
We start with the models with universal extra dimension, then we consider
the models with 
split fermions, and finally we deal with scenario of warped extra dimension. 
We discuss possible new sources of flavor and CP violations 
that these models could have beyond those in the CKM. 
We also study the impact of these new sources on the associated CP violating 
asymmetries to the rare $B$ decays. We show that in the case of universal
extra dimension, 
the flavor and the CP violation are given by the CKM matrix as in the SM
and no significant deviations
in the results of the CP asymmetries of the $B$ decays are expected. We
also emphasize that with 
split fermions and models with warped extra dimensions, where fermions and
gauge fields live in the bulk, 
the KK contributions to the $B_d-\bar{B}_d$ and $B_d \to \phi K_S$ occur at the 
tree level. The $\Delta M_{B_d}$ 
experimental limit then implies that the compactification scale is of
order $10^4$ GeV.   
With such a large scale, we find that only the split fermion models can
lead to a significant deviation 
between the CP asymmetries $S_{\phi K_S}$ and $S_{J/\psi K_S}$ and the
recent results by Belle can be accommodated.

The paper is organized as follows. In the next section, we present model
independent expressions
for the flavor violating interactions due to the KK-excitations of the
gauge fields in models 
with large extra dimensions. In section 3, we study the effect of the new
flavor on the rare 
$B$-processes in the models with universal extra dimension. Section 4 is
devoted for analyzing the 
impact of the new sources of flavor in split fermions scenarios in
$B_d-\bar{B}_d$ mixing and CP 
asymmetry of $B_d \to \phi K_S$. In section 5, we carry the same analysis
in the models with warped 
extra dimension. Our conclusions are presented in section 6.

\section{\large{\bf{New source of flavor violation}}}
 We start our analysis by examining the impact of the infinite towers of
the Kaluza-Klein (KK)
modes that occur in extra dimensions on the $\Delta B=2$ processes
($B$ here stands not for baryon number but the $b$ quark number); in
particular we study these effects for $B_d-\bar{B}_d$ 
mixing and the CP asymmetry $S_{J/\psi K_S}$, and the $\Delta B=1$
processes, like $B_d \to 
\phi K_S$.  Generally $\Delta M_{B_d}$ and $S_{J/\psi K_S}$ can be 
calculated via
\bea
\Delta M_{B_d}&=& 2 \vert \langle B_d \vert H_{\mathrm{eff}}^{\Delta B=2} \vert
\bar{B}_d \rangle \vert ,\nonumber \\
S_{J/\psi K_S}&=& \sin 2 \beta_{\mathrm{eff}},~~ \mathrm{where}~ 
\beta_{\mathrm{eff}}= \frac{1}{2} \mathrm{arg}\langle B_d \vert
H_{\mathrm{eff}}^{
\Delta B=2} \vert \bar{B}_d \rangle ,
\eea
where $H_{\mathrm{eff}}^{\Delta B=2}$ is the effective Hamiltonian for the
transition $\Delta B=2$. In the framework of the SM, 
\be
S^{SM}_{J/\psi K_S}= \sin 2 \beta, ~~~ \beta = \mathrm{arg}\left(-
\frac{(V_{CKM})_{cd} (V^*_{CKM})_{cb}}{
(V_{CKM})_{td} (V^*_{CKM})_{tb}} \right).
\ee
The effect of the KK contribution can be described by a dimensionless
parameter  $r_{KK}^2$ and a phase $2 \theta_{KK}$ defined as
\be
r_{KK}^2 ~ e^{2 i \theta_{KK}} = \frac{M_{12}(B_d)}{M_{12}^{\mathrm{SM}}(B_d)},
\ee
where $M_{12}(B_d) = M_{12}^{SM}(B_d) + M_{12}^{KK}(B_d)$, and
$M_{12}^{KK}$ is the new contribution arising from the exchange of
KK-modes. Thus $\Delta M_{B_d}$ is given by 
$\Delta M_{B_d} = 2 \vert M_{12}^{\mathrm{SM}}(B_d) \vert~ r_{KK}^2$
and the CP asymmetry of $B_d \to J/\psi K_S$ is given as
\be
S_{J/\psi K_S}= \sin 2 \beta_{\mathrm{eff}} = \sin (2 \beta + 2 \theta_{KK}),
\ee 
where $ 2 \theta_{KK} = \mathrm{arg}\left(1+
M_{12}^{KK}/M_{12}^{SM}\right)$. The SM contribution 
is known at NLO accuracy in QCD and it is given by
\be
M_{12}^{SM} = \left(\frac{G_F}{4\pi}\right)^2 M_W^2 \left(V_{td}
V^*_{tb}\right)^2 S_0(x_t)
\eta_B \left[\alpha_s(\mu_b)\right]^{-6/23} \left[1+
\frac{\alpha_s(\mu_b)}{4\pi} 
J_5 \right]\left(\frac{4}{3} m_{B_d} f_{B_d}^2 B_1(\mu) \right),
\ee
where $x_t= (m_t/m_W)^2$. The renormalization group evolution factors
$\eta_B$ and 
$J_5$ are given by $\eta_B= 0.551$ and $J_5=1.627$. The Inami-Lim
function $S_0(x)$ 
is given by
\be 
S_0(x) = \frac{4 x -11 x^2 + x^3}{4(1-x)^2} - \frac{3 x^3 \ln x}{2(1-x)^3}.
\ee

Now we turn to the effect of the KK contributions to the CP asymmetries of the 
$B_d\to \phi K_S$. The direct and the mixing CP asymmetry are respectively
given by \cite{KK-phiks}
\begin{equation}
C_{\phi K_S}=\frac{|\overline{\rho}(\phi K_S)|^2-1}{|\overline{\rho}(\phi
K_S)|^2+1}, 
\ \ 
S_{\phi K_S}=\frac{2Im \left[\frac{q}{p}~\overline{\rho}(\phi K_S)\right]} 
{|\overline{\rho}(\phi K_S)|^2+1},
\end{equation}
where $\overline{\rho}(\phi K_S)=\frac{\overline{A}(\phi K_S)}{A(\phi
K_S)}$. The
$\overline{A}(\phi K_S)$ and $A(\phi K_S)$ are the decay amplitudes of 
$\overline{B}_d^0$ and $B^0_d$  mesons, which can be written as
\begin{equation}
\overline{A}(\phi K_S)=\langle \phi K_S| \mathcal{H}_{\mbox{eff}}^{\Delta
B=1}|\overline{B}^0\rangle,  \ \ \ 
A(\phi K_S)=\langle \phi K_S| (\mathcal{H}_{\mbox{eff}}^{\Delta
B=1})^{\dagger}|B^0\rangle. 
\end{equation} 
The mixing parameters $q/p$ is given by
\begin{equation}
\frac{q}{p}= 
\sqrt{\frac{M_{12}^*-\frac{i}{2}\Gamma_{12}^*}{M_{12}-\frac{i}{2}\Gamma_{12}}},
\end{equation}
where 
\begin{equation}
\langle B^0|(\mathcal{H}^{\Delta
B=2}_{\mbox{eff}})^{\dagger}|\overline{B}^0\rangle \equiv
M_{12}-\frac{i}{2}\Gamma_{12}. 
\end{equation} 
The $\Delta B=1$ effective Hamiltonian, including the KK-mediation,
is given by
\be
\mathcal{H}_{\mbox{eff}}^{\Delta B=1} =  \sum_{i} (C^{SM}_i + C^{KK}_i) Q_i + 
(L \leftrightarrow R),
\ee
where $Q_i$ are the operators represent the $b \to s \bar{s} s$ transition. 
Thus, one obtains 
\be
A(\phi K_S) = A^{SM}(\phi K_S) + A^{KK}(\phi K_S).
\ee
In this respect and following the parametrization of the SM and the 
KK as in Ref.\cite{KK-phiks},
the CP asymmetries of $B_d \to \phi K_S$ are given by 
\begin{eqnarray}
S_{\phi K_S}&=&\Frac{\sin 2 \beta +2 R_{KK} \cos \delta_{KK} 
\sin(\theta'_{KK} + 2 \beta) +
R_{KK}^2 \sin (2 \theta'_{KK} + 2 \beta)}{1+ 2 R_{KK}
\cos \delta_{KK} \cos\theta'_{KK} +R_{KK}^2},
\label{sphi}\\
C_{\phi K_S}&=& - ~\Frac{2 R_{KK} \sin \delta_{KK} \sin \theta'_{KK}}
{1+ 2 R_{KK} \cos \delta_{KK} \cos \theta'_{KK} + R^2_{KK}} 
\end{eqnarray}
where $ R_{KK}= \vert A^{KK}/A^{SM}\vert$, $\theta'_{KK}=
\mathrm{arg}(A^{KK}/A^{SM})$, and $\delta_{KK}$ is the strong phase.
\section{\large{\bf{Flavor violation in universal extra dimension Scenario}}}
Recently there has been a growing interest concerning the models 
with one universal extra dimension (UED), as an alternative view of gauge
hierarchy problem of the SM \cite{Appelquist:2000nn}. In this class of
models, the compactification
scale is the only additional free parameter relative to the SM. Also, the
tree level KK contributions to low energy processes are absent, however,
their one loop contributions are important as have emphasized in Ref.
\cite{Appelquist:2000nn,Buras:2003mk}.
It was shown that for $1/R \lsim 400$ GeV, the KK impacts could be very 
significant. Nonetheless, the electroweak precision impose a lower bound
on $1/R$ to be larger than 300 GeV. 

Concerning the flavor violation and CP violation in this model, it is, 
as in the SM, given by the CKM matrix only. Therefore, one would not 
expect a significant deviation from the SM results in the CP asymmetries 
of the $B$ decays. In fact, within the UED scenario, the main effect of the KK 
modes on these processes is the modification of the Inami-Lim one loop
functions, as 
was found for other processes in Ref.\cite{Buras:2003mk}. We will show
that this 
modification is quite limited and can not explain the $2.7 \sigma$ deviation
from $\sin 2 \beta$ in the process $B_d \to \phi K_S$ announced by Belle 
and BaBar Collaborations \cite{Belle3,BaBar3}.
 
In the models with UED, the fifth dimension is 
compactified on the orbifold $S_1/Z_2$ to produce chiral fermion in 
four dimensions. There are infinite KK modes of the SM particle with 
universal masses
\be
m_{(n)}^2 = m_0^2 + \frac{n^2}{R^2},
\ee
where $m_0$ is the mass of the zero mode, which is the ordinary SM 
particles. It was noticed that the ratio $x_{i(n)} = m_{i(n)}^2/M^2_{W(n)}$
is a natural variable that enter the Inami-Lim functions \cite{Buras:2003mk}, where 
$m_{i(n)}$ are the masses of the fermionic KK modes and $m_{W(n)}$ are
the masses of the $W$ boson KK modes. The effective Hamiltonian for the 
$\Delta B=2$ transition in the UED is given by \cite{Buras:2003mk}
\be
\mathcal{H}_{\mathrm{eff}}^{\Delta B=2} = \frac{G_F^2}{16 \pi^2} M_W^2
(V_{tb}^* V_{td})^2 \eta_B S(x_t, 1/R) \left[\alpha_s^{(5)}(\mu_b)
\right]^{-6/23} \left[ 1+ \frac{\alpha_s^{(5)}(\mu_b)}{4\pi} J_5\right] Q\left(\Delta B=2
\right) +h.c.
\ee
where $S(x_t,1/R)$ is defined as
\be
S(x_t, 1/R)= S_0(x_t) + \sum_{n=1}^{\infty} S_{n}(x_t, x_n).
\ee 
The KK contributions are represented by the function $S_{n}(x_t, x_n)$.
A lower bound on the compactification scale $1/R \gsim 165$ GeV has been
obtained by requiring the KK contribution does not exceed the experimental
central value $\Delta M_{B_d} < 0.484~ (ps)^{-1}$ \cite{katri}.

It is worth mentioning that with the above expression of 
$\mathcal{H}_{\mathrm{eff}}^{\Delta B=2}$, the phase $\theta_{KK}$ 
defined in the previous section is identically zero, hence there is no 
new source of CP violation in this class of models and the CP asymmetry 
of $B_d \to J/\psi K_S$ would be given by the SM value, $\sin 2 \beta$, 
which is consistent with the experimental measurements.

Now we analyze the KK contributions to the CP asymmetry $B_d \to \phi K_S$ 
decay in this class of models with UED. The effective Hamiltonian for the 
$\Delta B=1$ transitions through the dominant gluon and chromomagnetic 
penguins is given by
\be
\mathcal{H}_{\mathrm{eff}}^{\Delta B=1} = - \frac{G_F}{\sqrt{2}} 
V_{tb} V_{ts}^* \left[ \sum_{i=3}^6 C_i O_i + C_g O_g + 
\left( L \leftrightarrow R \right) \right],
\ee
where $O_i$ and $O_g$ are the relevant local operators, which are given in 
Ref.\cite{Buchalla:1995vs}. The corresponding Wilson
coefficients are given as follows:
\be 
C_3(M_W) = -3 C_4(M_W) = C_5(M_W) = -3 C_6(M_W)= - \frac{\alpha_s}{24 \pi}
\tilde{E}(x_t,1/R),
\ee
and
\be
C_g(M_W) = - \frac{1}{2} E'(x_t , 1/R).
\ee
The functions $\tilde{E}(x_t,1/R)$ and $E'(x_t,1/R)$ are given by
$$ F(x_t, 1/R) = F_0(x_t) + \sum_{n=1}^{\infty} F_n(x_t,x_n),~~~~~ F\equiv
\tilde{E}, E',$$ 
where $\tilde{E}_{0,n}$ and $E'_{0,n}$ are given in Ref.\cite{Buras:2003mk}.
Taking into account the lower bound obtained on the compactification scale
from $\Delta M_{B_d}$ measurements, one finds that the ratio 
$R_{KK} \lsim 0.05$ and the phase $\theta'_{KK} \sim -0.02$. Clearly these
values are much smaller than the required values mentioned in Ref.\cite{KK-phiks}
in order to deviate $S_{\phi K_S}$ from $\sin 2 \beta$. Indeed we find that, 
in this case, the total $S_{\phi K_S}$ is given by $S_{\phi K_S} \simeq
0.72$. Thus, one can conclude that an experimental confirmation of a 
deviation of $S_{\phi K_S}$ from $\sin 2 \beta$ will disfavor models 
with UED.   

We point out that the original UED models do not address the neutrino mass
problem and an extended version of this model based on the gauge groups
$SU(2)_L \times U(1)_{I_{3R}}\times U(1)_{B-L}$ and $SU(2)_L \times
SU(2)_{R} \times U(1)_{B-L}$ which solves the neutrino
mass problem has been discussed in \cite{perez}. In the four
dimensional left-right models \cite{moh}, there is the well known $W_L-W_R$
exchange box graph \cite{beall} which makes a large contribution to flavor
changing hadronic processes such as $K-\bar{K}$ mixing. One might
therefore suspect that there will be similar contributions in the 5D
left-right model which will then lead to tighter constraints on
the compactfication scale. However it turns out that since the lightest 
$W_R$ mode in 5D LR models of the type discussed in ref.\cite{perez} is a KK mode with a
specific $Z_2\times Z'_2$ quantum number (+,-),
it only connects the known quarks and leptons (which
are zero modes of the 5D fermion field) to KK excitations of the
quarks with $Z_2\times Z'_2$ quantum numbers (+,-). Therefore there is no
box graph with $W_R$ and $W_L$ exchange that contributes to FCNC
processes. So only flavor changing diagrams
arise from exchange of KK modes of two $W_L$'s in a manner very similar to
the UED models with $SU(2)_L\times U(1)_Y$ gauge group.
This implies that the standard model 
prediction for $B_d \rightarrow \phi K_S$ CP asymmetry is unaffected in the
5D LR models. Furthermore, there are no $W_L-W_R$ mixing contributions to
penguin type graphs since the same $Z_2\times Z'_2$ symmetry forbids 
$W_L-W_R$ mixing. This is to be contrasted with the 4-dimensional
left-right model case, where the $W_L-W_R$ loop gluon penguin is indeed
present and gives a large
new CP violating contribution to $B_d \rightarrow \phi K_S$ and lead to
significant deviation from the standard model predictions \cite{raidal}.

Thus our first conclusion is that in all the UED type models constructed
so far, no significant deviation from the standard model prediction for 
$B_d \rightarrow \phi K_S$ CP asymmetry is to be expected.

\section{\large{\bf{Flavor violation in split fermions scenario}}}
Another class of extra dimensions which have drawn a lot of attention is 
the so called split fermion scenario \cite{Arkani-Hamed:1999dc}. 
The idea of split fermion is that  having different fermion localization
along the fifth direction may provide a geometrical way to look at the
fermion mass hierarchy. As in the UED type models, in these models, all
particles are considered to be in all five dimensions; but in contrast
with the UED models, the SM fermions $\psi_i$ are localized at 
different points in the fifth dimension with Gaussain profiles 
$\psi_i \sim e^{-(y-y_i)^2/\sigma^2}$. 
Here $y_i$ is the position of the quark in the fifth dimension 
and $\sigma$ is the width of the fermion wave functions with $\sigma \ll R$. 
In order to avoid introducing another hierarchy, one takes 
$\sigma/R \sim 0.1$. 
The quark mass matrices arise from the interaction
of fermions and the vacuum expectation value (vev) of Higgs zero mode and
 are given by
\bea
(M_u)_{ij} = \frac{v_0 (\lambda_u)_{ij}}{\sqrt{2}}
e^{\frac{-(\Delta^u_{ij})^2}{4\sigma^2}},\nonumber\\
(M_d)_{ij} = \frac{v_0 (\lambda_d)_{ij}}{\sqrt{2}}
e^{\frac{-(\Delta^d_{ij})^2}{4 \sigma^2}},
\eea 
where $\Delta_{ij} = \vert y_i - y_j \vert$ is the distance between
flavor $i$ and $j$.
The parameters $(\lambda_{u,d})_{ij}$ are the 5D Yukawa couplings, which are 
in general arbitrary matrices. In order to relate the hierarchy 
of the fermion masses to the locations of different fermion families, one
assumes that these couplings are of order unity. However, in order to 
avoid the factorizable form of the quark mass matrices which has always
two vanishing
eigenvalues, the $\lambda_{u,d}$ matrices can not be unit matrices. 
Thus, in this class of models the number of free parameters is larger 
than the number of the observed fermion masses and 
mixings and it is very easy to accommodate different types of Yukawa
textures with hierarchical or non-hierarchical features. Also with complex
$\lambda_{ij}$ one can get the SM phase in the CKM of order $\pi/2$.

Examples of hierarchical Yukawa couplings have been obtained in Ref. 
\cite{Grossman:2002pb,Mirabelli:1999ks,Chang:2002ww}, 
which fit all the quark and lepton masses and mixing angles. 
For instance, the solution of Ref.\cite{Mirabelli:1999ks} leads to the 
following position for the up and down quarks in the fifth dimension:
\be
y_{Q_L} \sim \sigma \left(\begin{array}{c}
		0\\
		14.2349\\
		8.20333 
\end{array} 
\right),~~~~ 	
y_{d_R} \sim \sigma \left(\begin{array}{c}
		19.4523\\
		5.15818\\
		10.1992 
\end{array} 
\right),~~~~ 	
y_{u_R} \sim \sigma \left(\begin{array}{c}
		6.13244\\
		20.092\\
		9.64483 
\end{array} 
\right), 	
\label{yi}
\ee
As we will explain below, in these models and due to the non-universal 
couplings with KK-gluon, both left- and right-handed rotations 
$V_{L,R}$ that diagonalize the mass matrix are observable. 
This is unlike the case in the SM where only $V_L$ rotations are physical
($V_{CKM} = V_L^{d^+} V_L^u$) and $V_R$ is completely decoupled. In general, 
$V_R$ matrix has six phases, these new phases 
might play an important rule in the CP violation of the $B$ system and 
this what we are going to examine below. It is also worth
mentioning that for a hierarchical $V_R$, this new KK effect is suppressed,
and therefore, in our analysis we choose the parameters $\lambda_{ij}$
such that $V_R$ is a non-hierarchical matrix.

The fields living in the bulk can be 
defined to be even or odd under the $Z_2$ parity. Thus we
will assume that the gauge boson in 5D bulk are even under the $Z_2$ 
and have the following Fourier expansion
\be
A_{\mu}(x,y) = \frac{1}{\sqrt{R}} A_{\mu}^{(0)} + \sqrt{\frac{2}{R}} 
\sum_{n=1}^{\infty} \cos \frac{n y}{R} A_{\mu}^{(n)} (x).
\ee
where $R$ is the size of the extra dimension. The relevant 
terms of the effective 4D Lagrangian as far as flavor is concerned are
given by
\bea
\mathcal{L} &=& \bar{d}_R M_d^{\mathrm{diag}} d_L +  
\bar{u}_R M_u^{\mathrm{diag}} u_L + \frac{g}{\sqrt{2}} W_{\mu} \bar{u}_L
\gamma^{\mu} V_{CKM} d_L \nonumber\\
&+& \sum_{n=1}^{\infty} \left[ \sqrt{2} g_s G_{\mu}^{A(n)} \left(\bar{d}_L \gamma^{\mu}
T^A U^{d(n)}_L d_L + \bar{d}_R \gamma^{\mu} T^A U^{d(n)}_R d_R \right) + (d
\leftrightarrow u)\right] + h.c.
\label{4dlag}
\eea
In obtaining Eq.(\ref{4dlag}), the fifth dimension has been integrated out and 
due to the Gaussain nature of the quark wave functions, only the interaction 
of the KK-gluon, 
$G_{\mu}^{A(n)}$, with zero mode mass eigenstate quarks at the points
$y_i$, given in Eq.(\ref{yi}), 
are picked. The matrices $U^{d(n)}_{L,R}$ are defined as  
\be
U^{d(n)}_{L(R)}= V^{d^\dagger}_{L(R)} C^{d(n)}_{L(R)} V^d_{L(R)}
\ee
where $V^d_{L,R}$ are the unitary matrices that diagonalize the down
quark mass matrix.
The non-universal couplings $C^{d(n)}_{L(R)}$ are given by
\be
C^{d(n)}_{L(R)}= \mathrm{diag}\left
\{\cos\left(n y_{d_{L(R)}}/R\right), 
\cos\left(n y_{s_{L(R)}}/R\right),\cos\left(n y_{b_{L(R)}}/R\right)\right\}.
\ee

The non-universality in the position of the different families at 
$y$-direction leads to a new source of flavor in $U_{L,R}^{q(n)}$ in addition 
to the usual $V_{CKM}$. It is clear that this flavor violation 
can be mediated at tree level by the KK modes of the gluon which makes 
it quite dangerous and strong bounds on the compactification scale have been 
obtained \cite{quiros}. 
Of course, there are similar contributions from the KK modes of the
$\gamma$, $Z$ and $W$ bosons; however, one expects that the largest effect 
is due to the KK gluon.
\begin{figure}[t]
\begin{center}
\hspace*{-7mm}
\epsfig{file=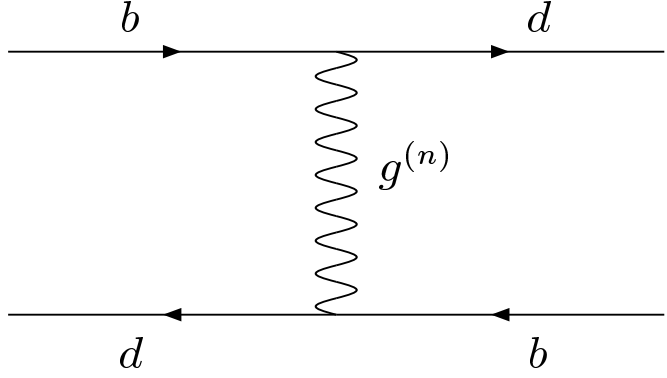,width=6cm,height=5cm}\\
\caption{{\small The KK-modes gluon contributions to the $B_d-\bar{B}_d$ mixing.}}
\label{khalil1}
\end{center}
\end{figure}

Unlike the SM and UED, the contribution from the KK-modes in split fermion 
models to the $B_d-\bar{B}_d$ mixing is at the tree level. The KK-gluon, 
shown in Fig.1, gives the dominant contributions to this process. However,
since the KK-modes of the gluino has no coupling between the $b$ and $c$ 
quarks at tree level, its contribution to the 
$B_d \to J/\psi K_S$ decay takes place at loop level. Therefore, the main
source of CP violation 
in this process is through the oscillation as in the SM. 

In the models with split fermions, the $\Delta B=2$ effective Lagrangian
for the KK-gluon exchange is given by
\be
\mathcal{L}^{\Delta B=2}_{KK} = \frac{2}{3}~ g_s^2~ \sum_{n=1}^{\infty}~
\frac{1}{M_n^2}~ \sum_{i,j=L,R}~
U^{d(n)^*}_{i(bd)}~ U^{d(n)}_{j(db)}~ (\bar{b}_i \gamma^{\mu}
d_{i})(\bar{b}_j \gamma^{\mu} d_{j}).
\ee
The hadronic matrix elements $\langle \bar{B}_d\vert (\bar{b}_i \gamma^{\mu}
d_{i}) (\bar{b}_j \gamma^{\mu} d_{j})\vert B_d \rangle$,
$i,j = L,R$ are given by 
\bea
&&\langle \bar{B}_d\vert  (\bar{b}_L \gamma^{\mu} d_{L})(\bar{b}_L
\gamma^{\mu} d_{L})\vert B_d \rangle
=\frac{1}{3} m_{B_d} f^2_{B_d} B_1(\mu),\\
&& \langle \bar{B}_d\vert  (\bar{b}_L \gamma^{\mu} d_{L})(\bar{b}_R
\gamma^{\mu} d_{R})\vert B_d \rangle 
= \frac{1}{4} \left(\frac{m_{B_d}}{m_b(\mu)+m_d(\mu)}\right)^2 m_{B_d}
f^2_{B_d} B_4(\mu),
\eea
where $m_{B_d}$ is the mass of the $B_d$ meason and $m_b$ and $m_d$ are
the masses of masses
of the $b$ and $d$ quarks at the scale $\mu$. In our analysis, we assume
that $m_b(m_b)=4.6$
GeV and $m_d(m_b)=0.0054$ GeV. The B-parameters are given by: $B_1(m_b) =
0.87$ and 
$B_4(m_b) = 1.16$. The other operators which obtained by exchanging
$L\leftrightarrow R$ from the 
above ones, have the same matrix elements, since the strong interaction
preserve parity. Thus, 
$M_{12}^{KK} = \langle \bar{B}_d\vert \mathcal{H}^{\Delta B=2}_{eff}
\vert B_d \rangle$ can be 
written as 
\bea
M_{12}^{KK} &=& \frac{2}{3} \frac{g_s^2}{M_c^2}\sum_{k,m=1}^3 \sum_{i,j=L,R}
\langle \bar{B}_d\vert (\bar{b}_i \gamma^{\mu}
d_{i}) (\bar{b}_j \gamma^{\mu} d_{j})\vert B_d \rangle \nonumber\\
&& (V_i^d)_{k3} (V_i^d)^*_{k1} 
(V_j^d)^*_{m1} (V_j^d)_{m3} \sum_{n}
\frac{\cos(\frac{n(y_i)_k}{R}) \cos(\frac{n(u_j)_m}{R})}{n^2}.
\label{m12kk}
\eea

For the example of 5D split fermion discussed above, 
the experimental limit $\Delta M_{B_d} < 3.2 \times 10^{-13}$ GeV leads
to a lower bound on the 
compactification scale of order $10^4$ GeV. As can be seen from table 1,
the bounds on $M_S$, for $\sigma/R=0.1, 0.05, 0.01$, derived from the
experimental measurement of
$\Delta M_{K} \simeq 3.5 \times 10^{-15}$ GeV are about one order of magnitude 
larger than those obtained from the experimental limit on $\Delta M_{B_d}$.
However in the $K-\bar{K}$ mixing, there is a significant uncertainty due
to the 
 low QCD corrections which make this constraint is unreliable. Therefore,
we will follow the conservative approach and will consider, through our
analysis, the $B_d-\bar{B}_d$ mixing
to constrain the compactification scale. 

\begin{table}[h]
\begin{center}
\begin{tabular}{|c||c|c|}
\hline
$\sigma/R$ & Lower bounds on $M_S$ from Exp. $\Delta M_{B_d}$  & Lower
bounds on $M_S$ from 
Exp. $\Delta M_K$
\\ \hline
\hline
0.1 & $10^4$ & $8 \times 10^5$  \\
\hline
0.05 & $8 \times 10^3$ & $3 \times 10^5$  \\
\hline
0.01 & $5 \times 10^3$ & $8 \times 10^4$  \\
\hline
\end{tabular}
\end{center}
\caption{The lower bounds on the comactification scale from the
experimental measurements of $\Delta M_{B_d}$
and $\Delta M_K$ as function of the parameter $\sigma/R$. The unit of the 
mass $M_S$ is in GeV.}
\label{table1}
\end{table}
Applying these constraints on the compactification scale, one finds that
the values of the phase
$\theta_{KK}$ are very small ($\lsim 10^{-2}$),  
hence the CP asymmetry $S_{J/\psi K_S}$ remains equal to the SM
prediction.
It is worth stressing that, the KK contribution to
$B_d-\bar{B}_d$ mixing (\ref{m12kk}) is proportional to 
the transition factor between the first and third generations which is
typically very small.
This counts as an extra suppression factor for the phase $\theta_{KK}$,
in addition to the large 
compactification scale. Thus, the chance of having a significant effect
on the CP asymmetry  $S_{J/\psi K_S}$ is reduced. 

\begin{figure}[t]
\begin{center}
\hspace*{-7mm}
\epsfig{file=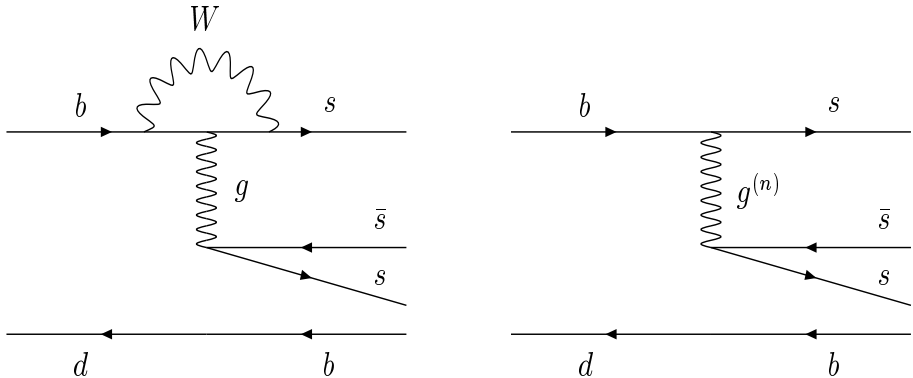,width=11cm,height=5cm}\\
\caption{{\small The SM and the KK-modes gluon contributions to the $B_d
\to \phi K_S$ decay.}}
\label{khalil2}
\end{center}
\end{figure}

Now we consider the KK contribution to the CP asymmetry $S_{\phi K_S}$. 
The tree level KK-modes gluon
contributions to the $B_d \to \phi K_S$ decay is shown in Fig. 2.  
The decay amplitude 
of $\bar{B}_d^0 \to \phi K_S$ (employing the naive factorization
approximation)  is given by
\begin{eqnarray}
\bar{A}(\phi K_S) &=& \langle \phi K_S \vert H_{\mathrm{eff}}^{\Delta
B=1} \vert \bar{B}_d^0 \rangle
\nonumber\\
&=& \frac{2}{3} g_s^2 \sum_{n=1}^{\infty} \frac{1}{M_n^2} \sum_{i,j= L,R} 
U^{d(n)^*}_{i(bs)} U^{(d(n)}_{j(ss)}  
\langle \phi K_S \vert (\bar{b}_i \gamma^{\mu} s_i)(\bar{s}_j \gamma_{\mu} s_j) 
\vert \bar{B}_d^0 \rangle .
\label{Aphik}
\end{eqnarray}
The matrix elements $\langle \phi K_S \vert (\bar{b}_i \gamma^{\mu}
s_i)(\bar{s}_j \gamma_{\mu} s_j) 
\vert \bar{B}_d^0 \rangle$ can be found in Ref.\cite{KK-phiks}. It is
remarkable that the coupling
$ U^d_{(bs)} U^d_{(ss)}$ in $\bar{A}(\phi K_S)$ is typically larger than
the coupling 
$ U^d_{(bd)} U^d_{(db)}$ of the $B_d-\bar{B}_d$ mixing, with one or two
orders of magnitude.
The size of this deviation strongly depends on the non-universality among
the parameters 
$C^{d(n)}_{L,R}$ and also on the flavor structure of the rotational
matrices $V^d_{L,R}$ 
that diagonalize the down quark mass matrix. 
This implies that it is quite natural to have significant $KK$
contributions to the CP asymmetry
of $B_d \to \phi K_S$ process and negligible one to the CP asymmetry of
$B_d \to J/\psi K_S$.

As shown in Eq.(\ref{sphi}), the 
deviation of $S_{\phi K_S}$ from $\sin 2 \beta$ is governed by the size
of the ratio $R_{KK}$ 
and the phase $\theta'_{KK}$. In this class of models, there
is a new source of flavor
and CP violation due to the impact of the right handed rotation $V^d_R$. 
We find that the size of the $R_{KK}$ and 
$\theta'_{KK}$ strongly depend on the favor structure of $V^d_R$ and 
the non-hierarchical form of $V_R^d$ is favored
to enhance their size and hence increase deviation between $S_{\phi K_S}$ and 
$S_{J/\psi K_S}$. It is also worth noting that the required form of 
$V_R^d$ can be obtained by tuning the arbitrary
5D Yukawa couplings $\lambda^d_{ij}$. We have explicitly studied different 
examples and found that $R_{KK}$ can vary from $\mathcal{O}(0.01)$ with 
$V^d_R \sim \mathcal{O}(V_{CKM})$ to $R_{KK} \sim 0.3$ for non-hierarchical   
$V_R^d$. Moreover, the phase $\theta'_{KK}$ is also quite sensitive to the structure 
of $V_R^d$ and it could be of order $\mathcal{O}(1)$. 

\begin{figure}[t]
\begin{center}
\hspace*{-7mm}
\epsfig{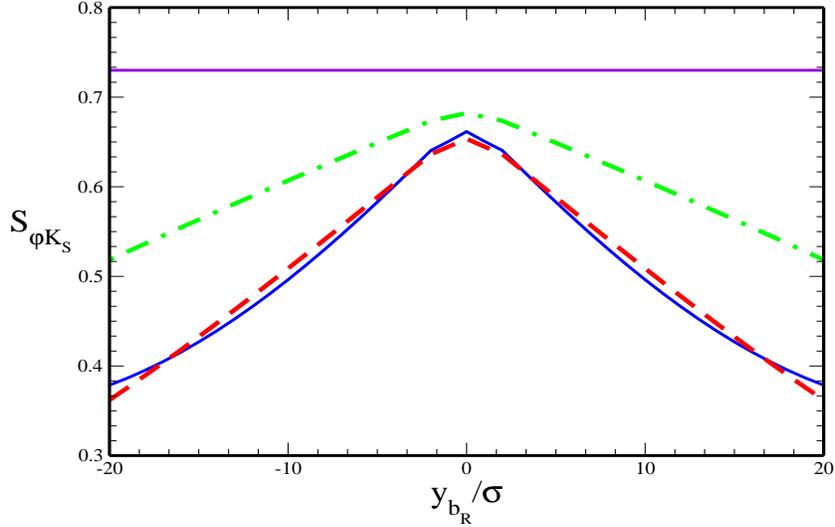}\\
\caption{{\small The figure shows the predictions for $S_{\phi K_S}$ as 
a function of $y_{b_R}/\sigma$ for various choice of the ratio
$\sigma/R$. The various values of $\sigma /R$ are $0.1$ for the solid
line, $0.05$ for dashed line and $0.01$ for dash-dotted line.}}
\label{kmfig3} \end{center}
\end{figure}

In Fig.3 we present the predicted values of the CP asymmetry of $S_{\phi K_S}$ as
function of the position of the right-handed bottom quark in the fifth dimension 
(which is one of the relevant parameters for $S_{\phi K_S}$). The other quark positions
have been fixed as in Eq.(\ref{yi}). Also the matrix $V_R$ has been chosen to be 
non-hierarchical and three values of $\sigma /R$ have used as in Table 1. 
As can be seen from this figure, it is possible to deviate $S_{\phi K_S}$ significantly
from $S_{J/\psi K_S}$. Moreover, with a larger mixing in $V_R$ (specially between the $2^{nd}$ and $3^{rd}$ generations) or with different set of positions from those 
given in Eq.(\ref{yi}), one gets smaller and even negative values for $S_{\phi K_S}$.

Thus in the split 
fermions scenario, the free parameters can be easily adjusted in order
to accommodate the anomaly $S_{\phi K_S}$ and $\sin 2 \beta$.

\section{\large{\bf{Flavor violation in warped extra dimensions}}}
Finally we consider the five dimensional models with warped geometry where
the SM fermions and gauge bosons correspond to bulk fields
\cite{warped}. The warped geometry 
has been proposed as solution of the hierarchy problem. In the original
model, the 
SM fields were localized to one of the boundaries and gravity is allowed
to propagate
in the bulk. However, it was realized that the scenarios of SM gauge
bosons and fermions
in the bulk may lead to a new flavor  and possible geometrical
interpretation for 
the hierarchy of quark and lepton masses. The Higgs field has to be
confined to the 
TeV brane in order to obtain the observable masses of the $W$ and $Z$
gauge bosons. 

We will consider the scenario of Ref.\cite{extra3}, based on the metric 
\be
ds^2 = e^{-2\sigma(y)} \eta_{\mu \nu} dx^{\mu} dx^{\nu} + d y^2,
\ee
where $\sigma(y)=\kappa \vert y \vert$ with $\kappa \sim M_P$ is the
curvature scale
determined by the negative cosmological constant in the five dimensional bulk.
The fermion fields reside in the bulk of this non-factorizable geometry
can be decomposed
as 
\be
\Psi(x,y) = \frac{1}{\sqrt{2 \pi R}} \sum_{n=0}^{\infty} \psi^{(n)}(x) 
e^{2 \sigma(y)} f_n(y).
\ee
Here $R$ is the radius of the compactified fifth dimension on an orbifold
$S_1/Z_2$
so that the bulk is a slice of $AdS_5$ space between two four dimensional
boundaries.
The zero mode wave function is given by \cite{Gherghetta:2000qt}
\be
f_0(y)=\frac{e^{-c \sigma(y)}}{N_0}.
\ee
where $c=m_{\psi}/\kappa$ and $m_{\psi}$ is the bulk mass term. Using the
orthonormal 
condition: $\frac{1}{2\pi R} \int_{-\pi R}^{\pi R} dy e^{\sigma}
f_0(y) f_0(y) =1 $, one finds 
that $N_0$ is given by
\be
N_0 = \sqrt{\Frac{e^{\pi \kappa R (1-2 c)} -1}{\pi \kappa R(1- 2 c)}}.
\ee
The tower of fermion KK excited states is not relevant to our discussion here. 
Also the massless gauge fields that propagate in this curved background
can be decomposed as 
\cite{Gherghetta:2000qt}
\be
A_{\mu}(x,y) = \frac{1}{\sqrt{2 \pi R}} \sum_{n=0}^{\infty} A_{\mu}^{(n)}(x) 
f_n^A(y),
\ee
with $f_n^A$ is given as
\be
f_n^A(y) = \frac{\sigma(y)}{N_n} \left[ J_1\left(\frac{m_n}{\kappa}
e^{\sigma(y)} \right) +
b^A(m_n) Y_1\left( \frac{m_n}{\kappa} e^{\sigma(y)}\right) \right],
\ee
where $J_1$ and $Y_1$ are the Bessel function of order one and 
$b^A(m_n)=-J_0(m_n/\kappa)/Y_0(m_n/\kappa)$. The coupling of the gauge KK
modes to  the fermion is given by 
\be
g_{ijn} = \frac{g^{(5)}}{(2\pi R)^{3/2}} \int_{-\pi R}^{\pi R} e^{\sigma(y)} f_i(y) f_j(y)
f_n^A(y) dy .
\ee
As in the split fermion scenario, the non-universal parameters $c_i$ lead
to non-universal 
couplings to the KK state of the gluon. In the basis of mass eigenstates
we have the following flavor dependent couplings
\be
U^{d(n)}_{L(R)} = V^{d^+}_{L(R)}~ g_{00n}~ V^d_{L(R)},
\ee
where the $g_{00n}$ is given by
\be
g_{00n} = g \left(\frac{1-2c}{e^{\pi \kappa R(1-2c)}
-1}\right) \frac{\kappa}{N_n} \int_0^{\pi R}
e^{(1-2c)\kappa y} \left[J_1\left(\frac{m_n}{\kappa} e^{\kappa y}\right) +
b^A(m_n) Y_1\left(\frac{m_n}{\kappa} e^{\kappa y}\right)\right].
\ee
The gauge coupling $g$ is defined as $g= g_5/\sqrt{2 \pi R}$, where $g_5$
is the $5D$ gauge 
coupling. Finally, the unitary matrices $V_{L,R}^d$ diagonalize the down quark
mass matrix which is 
given in this model as
\be
Y^d_{ij} = \frac{l_{ij}}{\pi \kappa R} f^d_{0iL}(\pi R) f^d_{0jR}(\pi R).
\ee
The dimensionless parameters $l_{ij}$ are defined as $\lambda_{ij}^{(5)}
\sqrt{\kappa}$
where $\lambda_{ij}^{(5)}$ are the $5D$ Yukawa couplings which are free
parameters to be fixed
by the observable masses and mixing. Therefore, this class of model with
warped geometry 
is similar to the models with split fermions on large extra dimensions
where the number 
of free parameters is larger than the number of the quark masses and
mixings. 
In this respect, any type of Yukawa textures can be obtained in this
models by tuning 
the free parameters $c_i$ and $l_{ij}$. For instance, in order to get the
non-hierarchical 
Yukawa textures that we have used in the previous section:
\begin{eqnarray}
\vert Y_d \vert \simeq \left( \begin{array}{ccc}
0.006 & 0.018 & 0.018\\
0.012 & 0.056 & 0.037\\
0.0189 & 0.005 & 0.99
\end{array}\right), ~~~~~~ \vert Y_u \vert \simeq \left( \begin{array}{ccc}
0.001 & 0.0017 & 0.019\\
0.0037 & 0.0069 & 0.0002\\
0.012 & 0.041 & 0.976
\end{array}\right),
\label{yuk-example}
\end{eqnarray}
one can start with any reasonable choice of $c_i$ which is consistent with
the observed hierarchy 
between different generations and between up and down type quarks and
find the corresponding 
couplings $\lambda^{(5)}_{ij}$ that lead to these textures. This choice
is not unique, for 
example, we can use the parameters $c_i$ listed in  Ref.\cite{huber}:
\begin{eqnarray}
&& c_{Q_1} = 0.643, ~~~~~~~~ c_{Q_2} = 0.583,~~~~~~~ c_{Q_3} = 0.317,\nonumber\\
&& c_{D_1} = 0.643, ~~~~~~~~ c_{D_2} = 0.601,~~~~~~~ c_{D_3} = 0.601,\\
&& c_{U_1} = 0.671, ~~~~~~~~ c_{U_2} = 0.528,~~~~~~~ c_{U_3} = -0.460.\nonumber
\end{eqnarray}
with the following values of $\lambda_{ij}^{(5)}$:
\begin{eqnarray}
\vert l^d_{ij} \vert \simeq \left( \begin{array}{ccc}
1.36 & 1.08 & 0.72\\
0.45 & 0.58 & 0.39\\
0.03 & 0.002 & 0.42
\end{array}\right),~~~~~~ \vert l^u_{ij} \vert =\left( \begin{array}{ccc}
28 & 0.8 & 0.66\\
17.7 & 0.57 & 0.0014\\
2.19 & 0.13 & 0.23
\end{array}\right)
\end{eqnarray}
to get the Yukawa couplings in Eq. (\ref{yuk-example}). 

As advocated above, this class of model leads to a flavor violation at
the tree level similar to 
the split fermion scenario. The effective Hamiltonians for the $\Delta
B=2$ and $\Delta B=1$ 
processes can be expressed as in Eqs. 26 and 30 but with
$U^{d(n)}_{L(R)}$ as given in Eq. 38. 
Having fixed the Yukawa couplings in both cases, thus the difference
between the 
flavor predictions of these two models is mainly due to the difference
between the 
non-universal couplings: $c^{d(n)}_{L,R}$ (in split fermions) 
and $g_{00n}$ (in warped geometry). In fact, it is easy to note that the
non-universality 
of $ c^{d(n)}_{L,R}$ is much stronger than the non-universality of
$g_{00n}$. For instance with 
our above assumption, the $g_{001}$ which gives the dominant contribution
is given by
$g_{001} \simeq g~ \mathrm{diag}\{ 0.04, 0.03, 4.7\times 10^{-7}\}$ while
$C^{d(1)}_{L}$ is 
given by $ C^{d(1)}_{L} \simeq \mathrm{diag}\{ 1, 0.14, 0.68 \}$.
Such large non-universality in split fermions scenario has been proved to
be very useful in order 
to get a significant effect on the CP asymmetry of $B_d \to \phi K_S$ process, 
even with compactification scale of order $10^4$ GeV. This makes the chance of 
saturated the new results of the CP asymmetry is possible within this
class of model.   

In warped geometry, the compactification scale is constrained by the
electroweak corrections to 
be $M_S \gsim 10$ TeV \cite{Hewett:2002fe}. It was shown that with this
value of $M_S$ most of 
the flavor violating processes induced by the KK excitations of the gauge 
boson are suppressed and quite below their experimental limits \cite{huber}. 
We have also verified that the KK contribution to $\Delta M_{B_d}$ is
quite negligible respect to the 
SM value. The same is true for the CP asymmetry $S_{J/\psi K_S}$ which is
essentially given by 
the SM value $\sin 2 \beta$.  Finally, we comment on the KK contributions
to $B_d \to \phi K_S$ in 
the warped extra dimensions. We find that the ratio of the KK amplitude
to the SM amplitude is 
very small ($R_{KK} \sim \mathcal{O}(10^{-2}$), which implies that it is
not possible in this 
class of models to deviate the value of $S_{\phi K_S}$ from the value of
$S_{J/\psi K_S}$. 
\section{Conclusion}
To summarize, we have searched for new flavor violating CP asymmetric 
effects in three different classes of extra
dimension models that could significantly alter the predictions of the
standard model and have found that the only models where measurable 
deviations can occur are those based on the split fermion hypothesis. In
other models such CP violating effects are suppressed because of the
constraints from the CP
conserving sector. 
While experimental confirmation of such deviations will not necessarily be an
evidence for such models, any lack of deviation from standard model will
impose constraints on the parameters of the model. For instance, agreement
with standard model would mean
that the Yukawa couplings in the split fermion models have specific
patterns. It may also impose constraints on the location of the flavors in
the fifth dimension in specific models. 

\section*{\bf \normalsize Acknowledgments}
The work of R. N. M. was supported by the US National
Science Foundation Grant No. PHY-0099544 and the work of S.K. was 
supported by PPARC. S.K would like to thank V. Sanz for  
enlightening discussion.
%

\end{document}